\documentclass[prd,aps,groupedaddress,showpacs]{revtex4}
\usepackage{epsf,epsfig,graphicx}
\newcounter{muni}

\begin{document}
\hbadness=10000 \pagenumbering{arabic}
%\rightline{Preprint-#}

\title{Penguin pollution in the $B^0\to J/\psi K_S$ decay}

\author{Hsiang-nan Li$^{1}$}
\email{hnli@phys.sinica.edu.tw}
\author{Satoshi Mishima$^2$}
\email{mishima@ias.edu}

\affiliation{$^{1}$Institute of Physics, Academia Sinica, Taipei,
Taiwan 115, Republic of China,} \affiliation{$^{1}$Department of
Physics, National Cheng-Kung University, Tainan, Taiwan 701,
Republic of China}

\affiliation{$^{2}$School of Natural Sciences, Institute for
Advanced Study, Princeton, NJ 08540, U.S.A.}

\begin{abstract}

We present the most complete analysis of the penguin correction to
the extraction of the standard-model parameter $\sin(2\phi_1)$
from the $B^0\to J/\psi K_S$ decay up to leading power in $1/m_b$
and to next-to-leading order in $\alpha_s$, $\phi_1$ being the
weak phase, $m_b$ the $b$ quark mass, and $\alpha_s$ the strong
coupling constant. The deviation $\Delta S_{J/\psi K_S}$ of the
mixing-induced CP asymmetry from $\sin(2\phi_1)$ and the direct CP
asymmetry $A_{J/\psi K_S}$ are both found to be of $O(10^{-3})$ in
a formalism that combines the QCD-improved factorization and
perturbative QCD approaches. The above results, different from
those of $O(10^{-4})$ and of $O(10^{-2})$ obtained in the previous
calculations, provide an important standard-model reference for
verifying new physics from the $B^0\to J/\psi K_S$ data.

\end{abstract}


\maketitle

The $B^0\to J/\psi K_S$ decay has been regarded as the golden mode
for extracting the standard-model parameter $\sin(2\phi_1)$
\cite{BS88}, $\phi_1$ being the weak phase of the
Cabbibo-Kobayashi-Maskawa (CKM) matrix element $V_{td}$. Though it
is believed that the penguin pollution in this mode is negligible,
a complete and reliable estimate of its effect is not yet
available. Such an estimate is essential, especially when $B$
physics is entering the era of precision measurement. Note that
the data of the mixing-induced CP asymmetries in $B$ meson decays,
such as of $S_{J/\psi K_S}$, are not included in the CKM
\cite{CKM} or UT \cite{UT} fitter. Understanding the penguin
effect, one will have a more concrete idea of whether a
discrepancy between the measured $S_{J/\psi K_S}$ and
$\sin(2\phi_1)$ from the fitters is a signal of new physics. At
the same time, the direct CP asymmetries of the $B\to J/\psi K$
decays, also related to the penguin correction, are expected to be
vanishingly small. A reliable estimate of the penguin effect, when
confronted with the precision measurement of $A_{CP}(B\to J/\psi
K)$, can reveal a new physics signal. It has been claimed that
$A_{CP}(B\to J/\psi K)$ observed at 1\% or higher would indicate
new physics definitely \cite{HNS06}.

The previous calculation of the deviation $\Delta S_{J/\psi
K_S}\equiv S_{J/\psi K_S}-\sin(2\phi_1)\sim O(10^{-4})$
\cite{HB04} has taken into account the corrections to the
$B$-$\bar B$ mixing \cite{JSH} and to the decay amplitude.
However, only the $u$-quark-loop contribution from the tree
operators was included in the latter. We shall point out that the
contribution from the penguin operators has been overlooked in
\cite{HB04}, which may result in a larger deviation. The
model-independent analysis in \cite{CPS05} gave $\Delta S_{J/\psi
K_S}=0.000\pm 0.012$, whose theoretical uncertainty is comparable
to the current systematical error in the $S_{J/\psi K_S}$
measurement. This method relies on the input of the $B^0\to
J/\psi\pi^0$ data in order to fix the penguin amplitude (including
its strong phase). The experimental error then propagates into
their uncertainty of $O(10^{-2})$, which should have been
overestimated. Though the $B^0\to J/\psi\pi^0$ data will get
precise in the future, SU(3) symmetry breaking effects are also
going to emerge, and one will lose the control of the theoretical
uncertainty eventually, if using this method. Hence, we are
motivated to reinvestigate this important subject by computing the
complete penguin correction to the $B^0\to J/\psi K_S$ amplitude.

In this work we shall adopt a special formalism \cite{CL05} that
combines the QCD-improved factorization (QCDF) \cite{BBNS} and
perturbative QCD (PQCD) \cite{KLS,LUY} approaches. The
factorizable contribution in the $B\to J/\psi K$ decays contains
the $B\to K$ form factor $F^{BK}_+(m_{J/\psi}^2)$ evaluated at the
invariant mass squared of the $J/\psi$ meson. Because of the heavy
$m_{J/\psi}$, the energy release involved in
$F^{BK}_+(m_{J/\psi}^2)$ is small, and it is unlikely to further
factorize it into a convolution of a hard kernel with the $B$
meson and kaon distribution amplitudes as in PQCD. Therefore, we
adopt the QCDF formalism to handle the factorizable amplitude. The
nonfactorizable spectator contribution is essential in the
color-suppressed category of $B$ meson decays \cite{YL,KKLL}, the
reason the naive factorization assumption \cite{BSW} does not
apply to the $B\to J/\psi K$ decays. For this piece, QCDF is not
appropriate due to the end-point singularity from vanishing parton
momenta, when the twits-3 kaon distribution amplitudes are
included \cite{Cheng}. Therefore, we employ the PQCD approach
based on $k_T$ factorization theorem, which is free of the
end-point singularity. As argued in \cite{CSLL}, the
nonfactorizable contribution has a characteristic hard scale
higher than that in the $B$ meson transition form factor. This
mixed formalism has been applied to $B$ meson decays into various
charmonium states \cite{CL05}, and results for the branching
ratios are consistent with the observed values.

The mixing-induced CP asymmetry $S_{J/\psi K_S}$ and the direct CP
asymmetry $A_{J/\psi K_S}$ of the $B^0\to J/\psi K_S$ decay are
defined by
\begin{eqnarray}
& &S_{J/\psi K_S}={2\,{\rm Im}\,\lambda_{J/\psi K_S} \over
1+|\lambda_{J/\psi K_S}|^2}\;,\;\;\;\; A_{J/\psi
K_S}={|\lambda_{J/\psi K_S}|^2-1 \over
1+|\lambda_{J/\psi K_S}|^2}\;,\nonumber\\
& & \lambda_{J/\psi K_S}=\frac{q}{p} {{\cal A}({\bar B}^0 \to
J/\psi K_S) \over{\cal A}(B^0 \to J/\psi K_S)} \;, \label{mix}
\end{eqnarray}
where the ratio $q/p$ is related to the non-diagonal elements of
the mixing matrix, and ${\cal A}$ the decay amplitude. Note that a
tiny $K_L$ admixture could exist in the experimentally
reconstructed $K_S$ final state, such that CP violation in the
$K$-$\bar K$ mixing contributes to $S_{J/\psi K_S}$ and $A_{J/\psi
K_S}$. However, we shall concentrate only on the corrections from
the $B$ meson system in this work, and refer the inclusion of the
corrections from the kaon system to \cite{GKL02}.

We aim at an accuracy up to leading power in $1/m_b$ and to
next-to-leading order in $\alpha_s$, $m_b$ being the $b$ quark
mass and $\alpha_s$ the strong coupling constant. At this level of
accuracy, the correction to the mixing factor $q/p$, computed
reliably in the framework of operator product expansion with the
large logarithm $\alpha_s\ln(m_W/m_b)$ being resummed \cite{HB04},
gives
\begin{eqnarray}
\Delta S_{J/\psi K_S}^{\rm mix}
=(2.08\pm 1.23)\times 10^{-4}\;,\label{cm}
\end{eqnarray}
where $m_W$ is the $W$ boson mass. Note that the old value
$\sin(2\phi_1)\approx 0.736$ has been input to derive the above
result. Using the updated one $\sin(2\phi_1)\approx 0.684$
\cite{Hazumi} changes Eq.~(\ref{cm}) by at most 10\%, namely, by
$O(10^{-5})$, which is absolutely negligible. The influence of the
$B$-$\bar B$ mixing on the tagging leads to the direct CP
asymmetry \cite{HB04,IB},
\begin{eqnarray}
A_{J/\psi K_S}^{\rm mix} = (2.59 \pm 1.48)\times
10^{-4}\;,\label{acp}
\end{eqnarray}
which is consistent with the estimate in \cite{LLNP}. Both
Eqs.~(\ref{cm}) and (\ref{acp}) are of $O(10^{-4})$.

The $B^0\to J/\psi K^0$ decay amplitude is decomposed into
\begin{eqnarray}
{\cal A}(B^0 \to J/\psi K^0) =V_{cb}^*V_{cs}\, \left({\cal
A}^{(c)}_{J/\psi K^0}+{\cal A}^{(t)}_{J/\psi K^0}\right) +
V_{ub}^*V_{us}\, \left({\cal A}^{(u)}_{J/\psi K^0}+{\cal
A}^{(t)}_{J/\psi K^0}\right)\;,
\end{eqnarray}
where the unitarity relation
$V_{tb}^*V_{ts}=-V_{ub}^*V_{us}-V_{cb}^*V_{cs}$ for the CKM matrix
elements has been inserted, and the second term carries the weak
phase $\phi_3$. The dominant tree amplitude ${\cal
A}^{(c)}_{J/\psi K^0}$ is responsible for the $B^0\to J/\psi K^0$
branching ratio, and ${\cal A}^{(u,t)}_{J/\psi K^0}$ are the
penguin pollution in the decay. The amplitude
${\cal A}^{(u)}_{J/\psi K^0}$, receiving the contribution from the
$u$-quark loop through the BSS mechanism \cite{BSS}, has been
estimated naively in \cite{HB04}, while ${\cal A}^{(t)}_{J/\psi
K^0}$ from the penguin operators has been missed. We shall
calculate these amplitudes in our formalism below.

The leading-power QCDF formulas for the factorizable contributions
to ${\cal A}^{(c,t)}_{J/\psi K^0}$ are written as \cite{CL05}
\begin{eqnarray}
{\cal A}^{(c)f}_{J/\psi K^0}&=& 2\sqrt{2N_c}\int_0^1
dx_3\Psi^L(x_3)a_2(x_3,t)\,
F_+^{BK}(m_{J/\psi}^2)\;,\nonumber\\
{\cal A}^{(t)f}_{J/\psi K^0}&=&2\sqrt{2N_c}\int_0^1
dx_3\Psi^L(x_3)\left[a_3(x_3,t)+a_5(x_3,t)\right]\,
F_+^{BK}(m_{J/\psi}^2) \;,\label{f}
\end{eqnarray}
with $N_c=3$ being the number of colors, $\Psi^L$ the twist-2
$J/\psi$ meson distribution amplitude, and $x_3$ the momentum
fraction carried by the $\bar c$ quark. The effective Wilson
coefficients including the $O(\alpha_s)$ vertex corrections
are given by
\begin{eqnarray}
a_2(x,\mu)&=&
C_1(\mu)+\frac{C_2(\mu)}{N_c}\left[1+\frac{\alpha_s(\mu)}{4\pi}C_F
\left(-18+12\ln\frac{m_b}{\mu}+f_I(x)\right)\right]
\;,\nonumber\\
a_3(x,\mu)&=&C_3(\mu)+C_9(\mu)+\frac{1}{N_c}
\left[C_4(\mu)+C_{10}(\mu)\right]\left[1
+\frac{\alpha_s(\mu)}{4\pi}C_F
\left(-18+12\ln\frac{m_b}{\mu}+f_I(x)\right)\right]
\;,\nonumber\\
a_5(x,\mu)&=&
C_5(\mu)+C_7(\mu)+\frac{1}{N_c}\left[C_6(\mu)+C_8(\mu)\right]
\left[1 +\frac{\alpha_s(\mu)}{4\pi}C_F
\left(6-12\ln\frac{m_b}{\mu}-f_I(x)\right)\right] \;,\label{a235}
\end{eqnarray}
where the explicit expression for the function $f_I$ can be found
in \cite{Cheng,chay},
\begin{eqnarray}
f_I(x)=\frac{3(1-2x)}{1-x}\ln x-3\pi
i+3\ln(1-r_3^2)+\frac{2r_3^2(1-x)}{1-r_3^2 x} \;,
\end{eqnarray}
with the ratio $r_3=m_{J/\psi}/m_B$, $m_B$ being the $B$ meson
mass. Compared to \cite{CL05}, the hard scale $t$ for the strong
coupling constant and for the Wilson coefficients has been chosen
as the larger internal momentum transfers between
$(P_2+x_3P_3)^2\approx x_3(m_B^2-m_{J/\psi}^2)$ and
$(P_2+\overline{x_3}P_3)^2\approx
\overline{x_3}(m_B^2-m_{J/\psi}^2)$ with the notation
$\overline{x_3}=1-x_3$, i.e.,
\begin{eqnarray}
t=\max\left(\sqrt{x_3(m_B^2-m_{J/\psi}^2)},
\sqrt{\overline{x_3}(m_B^2-m_{J/\psi}^2)}\ \right)\;.
\end{eqnarray}
This choice is more consistent with the derivation of the PQCD
formalism \cite{CL} and follows the well-known
Brodsky-Lepage-Mackenzie procedure \cite{BLM}.

The leading-power $O(\alpha_s)$ nonfactorizable spectator
amplitudes in PQCD are quoted from \cite{CL05}:
\begin{eqnarray}
{\cal A}^{(c)nf}_{J/\psi K^0}= {\cal M}(a_2^\prime)\;,\;\;\;\;
{\cal A}^{(t)nf}_{J/\psi K^0}={\cal M}(a_3^\prime-a_5^\prime) \;,
\label{na}
\end{eqnarray}
where the effective Wilson coefficients are defined by
\begin{eqnarray}
a_{2}^{\prime }(\mu) =\frac{C_{2}(\mu)}{N_{c}}\;,\;\;\;\;
a_{3}^{\prime }(\mu) =\frac{1}{N_{c}} \left[
C_{4}(\mu)+C_{10}(\mu)\right]\;,\;\;\;\; a_{5}^{\prime}(\mu)
=\frac{1}{N_{c}}\left[ C_{6}(\mu)+C_{8}(\mu)\right]\;.\label{nwc}
\end{eqnarray}
The amplitude ${\cal M}$ is written as
\begin{eqnarray}
{\cal M}(a_i^\prime) &=& 16\pi C_{F}\sqrt{2N_{c}} \int_{0}^{1}dx_1
dx_2 dx_3 \int_{0}^{\infty }b_{1}db_{1} \phi _{B}(x_{1},b_{1})
\nonumber \\
&& \times \Big\{ \Big[ (1-2r^{2}_{3})\overline{x_3}\, \phi^A_{K} (
\overline{x_2} )\Psi^{L}(x_{3})
+\frac{1}{2} r^{2}_{3}\, \phi^A_{K}(\overline{x_2})\Psi^{t }(x_{3})
\nonumber \\
&&\ \ \ \ \ - r_{2} (1-r^2_3)x_2\, \phi^{P}_{K}(\overline{x_2})
\Psi^{L}(x_{3})+ r_{2} \left(
2r^{2}_{3}\overline{x_3}+(1-r^2_3)x_2 \right)
\phi^{T}_{K}(\overline{x_2}) \Psi^{L}(x_{3})
  \Big]\nonumber \\
&&\ \ \ \times \alpha_s(t_d^{(1)})a'_{i}(t_d^{(1)})S(t_d^{(1)})\,
h_d^{(1)}(x_1,x_2,x_3,b_1)
\nonumber \\
%%%%
&&\ \ \  - \Big[ (x_3+(1-2r^2_{3})x_{2})\,\phi^A_{K} (\overline{x_2})
\Psi^{L}(x_{3})+r^{2}_{3}\, \bigl(2r_{2}\phi^{T}_{K}(\overline{x_2})
-\frac{1}{2}\phi^A_{K}(\overline{x_2})\bigr)\Psi^{t}(x_{3})
 \nonumber \\
&&\ \ \ \ \
 -r_{2} (1-r^2_3)x_2\, \phi^{P}_{K}(\overline{x_2}) \Psi^{L}(x_{3})- r_{2}
\left( 2r^{2}_{3}x_3+(1-r^2_3)x_2 \right) \phi^{T}_{K}(\overline{x_2})
\Psi^{L}(x_{3})\Big]\nonumber \\
&&\ \ \  \times \alpha_s(t_d^{(2)})a'_{i}(t_d^{(2)})S(t^{(2)}_d)\,
h_d^{(2)}(x_1,x_2,x_3,b_1) \Big\} \;,\label{psi4}
\end{eqnarray}
with the color factor $C_F=4/3$, the mass ratio $r_2=m_{0K}/m_B$,
$m_{0K}$ being the chiral enhancement scale associated with the
kaon, the impact parameter $b_1$ conjugate to the transverse
momentum of the spectator quark, the two-parton twist-3 $J/\psi$
meson distribution amplitude $\Psi^{t}$, and the notation
$\overline{x_2}=1-x_2$. The explicit expressions of the hard
functions $h_d^{(1,2)}$, the hard scales $t_d^{(1,2)}$, and the
Sudakov factor $S$ are referred to \cite{CL05}.

The penguin correction from the $u$-quark loop, which is of
next-to-leading order in $\alpha_s$, can be expressed in terms of
the effective Hamiltonian \cite{HB04},
\begin{eqnarray}
{\cal H}_{\rm eff}^{(u)}&=&-\frac{G_F}{\sqrt{2}}V_{us}^*V_{ub}
\left[\frac{\alpha}{3\pi}e_ue_c \left(N_cC_1+C_2\right)(\bar
cc)_V(\bar s b)_{V-A}+\frac{\alpha_s}{3\pi}C_2(\bar cT^ac)_V(\bar
sT^ab)_{V-A}\right]\nonumber\\
& &\times\left(\frac{2}{3}-\ln\frac{l^2}{\mu^2}+i\pi\right)\;.
\label{uq}
\end{eqnarray}
The first (second) operator arises from the process, where the
$c$-quark pair is produced through the photon (gluon) emission
from the $u$-quark loop. $l^2$ is the invariant mass squared of
the photon or gluon. Notice the constant $2/3$ different from
$5/3$ in \cite{HB04}, since we have considered the Fierz
transformation of the four-fermion operators with the
anti-commuting $\gamma_5$ in $D$ dimensions under the NDR scheme
\cite{LMS05}. We have also included the appropriate color factor
$N_c$ and the charge factor $e_ue_c=4/9$. It is easy to observe
that the first operator can be decomposed into $O_7$ and $O_9$,
and the second operator into $O_{3-6}$.

The leading contribution from the first operator is a factorizable
amplitude,
\begin{eqnarray}
{\cal A}^{(u1)f}_{J/\psi K^0}&=&2\sqrt{2N_c}\int_0^1
dx_3\Psi^L(x_3)a_{ew}(x_3,t)\, F_+^{BK}(m_{J/\psi}^2)
\;,\label{few}
\end{eqnarray}
with the effective Wilson coefficient,
\begin{eqnarray}
a_{ew}(x,\mu)=-\frac{\alpha}{3\pi}e_ue_c\left[N_cC_1(\mu)+C_2(\mu)\right]
\left(\frac{2}{3}-\ln\frac{m_{J/\psi}^2}{\mu^2}+i\pi\right)\;,\label{ew}
\end{eqnarray}
where $l^2$ has been set to $m_{J/\psi}^2$. We have examined the
average hard scale for the factorizable contribution, which is
slightly above 3 GeV (the average hard scale for the
nonfactorizable contribution is between 1 and 1.5 GeV). Simply
comparing Eq.~(\ref{ew}) with Eq.~(\ref{a235}) for $\mu\approx 3$
GeV, we find that the magnitude of Eq.~(\ref{few}) is less than
5\% of $|{\cal A}^{(t)f}_{J/\psi K^0}|$. Including the
higher-order terms, such as the nonfactorizable amplitude, the
effect from the first operator is expected to be around 5\%, and
can be safely dropped.

For the second operator in Eq.~(\ref{uq}) to contribute, an
additional gluon must attach the $c$-quark pair in the color-octet
state. If it is a soft gluon, the corresponding nonperturbative
input is the three-parton $J/\psi$ meson distribution amplitude,
defined via the nonlocal matrix element \cite{BB},
\begin{eqnarray}
\langle J/\psi(P,\epsilon^{*(\lambda)})|\bar
c(-z)g_s G_{\mu\nu}(vz)\gamma_\alpha c(z)|0\rangle &
&=if_{J/\psi}m_{J/\psi}P_{\alpha}\left[P_{\nu}\epsilon_{\perp\mu}^{*(\lambda)}-
\epsilon_{\perp\nu}^{*(\lambda)}P_{\mu}\right] {\tilde
V}(v,P\cdot z)\nonumber\\
& &+if_{J/\psi}m_{J/\psi}^3\frac{\epsilon^{*(\lambda)}\cdot
z}{P\cdot z} \left[P_{\mu}g_{\alpha\nu}^\perp-
P_{\nu}g_{\alpha\mu}^\perp\right] {\tilde
\Phi}(v,P\cdot z)\nonumber\\
& &+if_{J/\psi}m_{J/\psi}^3\frac{\epsilon^{*(\lambda)}\cdot
z}{(P\cdot z)^2} P_{\alpha}\left[P_{\mu}z_{\nu}-
P_{\nu}z_{\mu}\right] {\tilde \Psi}(v,P\cdot z)\;,\label{mat1}
\end{eqnarray}
with the polarization vectors $\epsilon^{(\lambda)}$ of the
$J/\psi$ meson, the gluon field strength tensor $G_{\mu\nu}$, and
the projector $g_{\mu\nu}^\perp=g_{\mu\nu}-(P_\mu z_\nu+P_\nu
z_\mu)/(P\cdot z)$, $z$ being the coordinate of the $\bar c$ quark
field. Because $m_{J/\psi}/m_B$ is not a small ratio, we extend
our investigation to twist 4. The twist-3 distribution amplitude
$\tilde V$, involved only in decays into transversely polarized
$J/\psi$ mesons, is irrelevant here. However, it could be a source
of uncertainty for the extraction of $\sin(2\phi_1)$ from the
$B\to J/\psi K^*$ decays. The twist-4 distribution amplitudes
$\Phi(x_c,x_{\bar c},x_g)$ and $\Psi(x_c,x_{\bar c},x_g)$, from
the Fourier transformation of $\tilde \Phi$ and $\tilde \Psi$,
respectively, are anti-symmetric under the exchange of the
momentum fraction $x_c$ of the $c$ quark and $x_{\bar c}$ of the
${\bar c}$ quark \cite{BB}, $x_g$ being the soft gluon momentum
fraction. Because the hard kernel associated with the $u$-quark
loop is symmetric under the exchange of $x_c$ and $x_{\bar c}$,
its convolution with $\Phi$ and $\Psi$ diminishes. That is, the
contribution from the second operator through higher-power terms
is suppressed.

If the additional gluon is hard, the resultant contribution, being
of $O(\alpha_s^2)$ (the gluon emitted by the $u$-quark loop is
also hard), is beyond the scope of this paper. Nevertheless, we
shall have a closer look at this piece. When the additional gluon
is emitted by the spectator quark, the corresponding amplitude has
the formula the same as ${\cal A}^{(t)nf}_{J/\psi K^0}$ but with
the replacement,
\begin{eqnarray}
a'_{3,5}\to \frac{\alpha_s}{12N_c\pi}C_2
\left(\frac{2}{3}-\ln\frac{l^2}{\mu^2}+i\pi\right)\;.
\label{rep}
\end{eqnarray}
Since the argument of ${\cal A}^{(t)nf}_{J/\psi K^0}$ depends on
the difference of $a'_3$ and $a'_5$ as shown in Eq.~(\ref{na}),
the above corrections cancel each other exactly. When the
additional gluon is emitted by the $b$ quark, the $s$ quark, or
the $u$-quark loop, the corresponding amplitude involves a
two-loop calculation, for which there is no simple estimation. The
two-loop vertex corrections from various operators are expected to
be finite, so their complete analysis for both ${\cal
A}^{(u)}_{J/\psi K^0}$ and ${\cal A}^{(t)}_{J/\psi K^0}$ may be
required in the future. Following the above analysis, we shall
also ignore the $c$-quark-loop correction, which modifies only the
prediction for the branching ratio slightly.

The choices of the $b$ quark mass, the $B$ meson wave function
$\phi_B$, the $B$ meson lifetime, the twist-2 kaon distribution
amplitude $\phi^A_K$, the two-parton twist-3 kaon distribution
amplitudes $\phi_K^P$ and $\phi_K^T$, the $B$ meson and kaon
decay constants, the chiral enhancement scale, and the CKM matrix
elements, including their allowed ranges, are the same as in
\cite{LM062}. We take $m_{J/\psi}=3.097$ GeV, and the $J/\psi$
meson distribution amplitudes \cite{CL05,BC04},
\begin{eqnarray}
\Psi^L(x)&=&N^L\,\frac{f_{J/\psi}}{2\sqrt{2N_c}}x(1-x)
\left[\frac{x(1-x)}{1-4\alpha x(1-x)}\right]^{\alpha}\;,\nonumber\\
\Psi^t(x)&=&N^t\,\frac{f_{J/\psi}}{2\sqrt{2N_c}}(1-2x)^2
\left[\frac{x(1-x)}{1-4\alpha x(1-x)}\right]^{\alpha}\;,
\end{eqnarray}
with the decay constant $f_{J/\psi}=405$ MeV \cite{Melic02}. The
shape parameter will be varied between $\alpha=0.7\pm 0.1$ to
acquire the theoretical uncertainty from these nonperturbative
inputs. It turns out that our results are not sensitive to the
variation of $\alpha$. The normalization constants $N^{L,t}$ are
determined by $\int dx\Psi^{L,t}(x)=f_{J/\psi}/(2\sqrt{2N_c})$,
giving the central values $N^L=9.58$ and $N^t=10.94$. The form
factor $F_+^{BK}(m_{J/\psi}^2)=0.62\pm 0.09$ is quoted from the
light-cone sum-rule calculation \cite{BZ04}, to which 15\%
theoretical uncertainty has been assigned.

It is observed that ${\cal A}^{(c)f}_{J/\psi K^0}$ and ${\cal
A}^{(c)nf}_{J/\psi K^0}$ are of the same order, confirming the
importance of the nonfactorizable contribution in color-suppressed
$B$ meson decays. Moreover, their real parts cancel, such that
${\cal A}^{(c)}_{J/\psi K^0}={\cal A}^{(c)f}_{J/\psi K^0}+{\cal
A}^{(c)nf}_{J/\psi K^0}$ is almost imaginary. The amplitudes
${\cal A}^{(c,t)}_{J/\psi K^0}$ lead to the branching ratio, and
$\Delta S_{J/\psi K_S}$ and $A_{J/\psi K_S}$ from the correction
to the decay,
\begin{eqnarray}
B(B^0\to J/\psi K^0) &=&
\left( 6.6^{+3.7\,(+3.7)}_{-2.3\,(-2.3)} \right) \times 10^{-4}
\;,\nonumber\\
\Delta S_{J/\psi K_S}^{\rm decay} &=&
\left( 7.2^{+2.4\,(+1.2)}_{-3.4\,(-1.1)} \right) \times 10^{-4}
\;,\nonumber\\
A_{J/\psi K_S}^{\rm decay} &=&
- \left( 16.7^{+6.6\,(+3.8)}_{-8.7\,(-4.1)} \right) \times 10^{-4}
\;,\label{decay}
\end{eqnarray}
where the errors in the parentheses arise only from the variation
of the hadronic parameters. The result for the $B^0\to J/\psi K^0$
branching ratio is in agreement with the data $B(B^0\to J/\psi
K^0)=(8.72\pm 0.33)\times 10^{-4}$ \cite{PDG}. A form factor
$F^{BK}(m_{J/\psi}^2)$ increased by 15\% and larger Gegenbauer
coefficients in the kaon distribution amplitudes can easily
account for the central value of the data. Our $\Delta S_{J/\psi
K_S}^{\rm decay}$ ($A_{J/\psi K_S}^{\rm decay}$) from the penguin
operators is about twice of the naive estimation from the
$u$-quark-loop contribution \cite{HB04}, and has an opposite (the
same) sign. Both $\Delta S_{J/\psi K_S}^{\rm decay}$ and
$A_{J/\psi K_S}^{\rm decay}$ in Eq.~(\ref{decay}) indicate the
$O(10^{-3})$ penguin pollution in the $B\to J/\psi K_S$ decay,
consistent with the conjecture made in \cite{GKL02}.

We have checked that the correction to the mixing does not modify
the denominator $1+|\lambda_{J/\psi K_S}|^2$ in the definitions of
$S_{J/\psi K_S}$ and $A_{J/\psi K_S}$ in Eq.~(\ref{mix}). Hence,
it is legitimate to simply add the values in Eqs.~(\ref{cm}) and
(\ref{acp}) to Eq.~(\ref{decay}). We then derive $\Delta S_{J/\psi
K_S}$ and $A_{J/\psi K_S}$ in the most complete analysis up to
leading-power in $1/m_b$ and to next-to-leading order in
$\alpha_s$,
\begin{eqnarray}
\Delta S_{J/\psi K_S}&=&\Delta S_{J/\psi K_S}^{\rm decay}+\Delta
S_{J/\psi K_S}^{\rm mix}=\left(
9.3^{+3.6}_{-4.6} \right) \times 10^{-4} \;,\nonumber\\
A_{J/\psi K_S}&=&A_{J/\psi K_S}^{\rm decay}+A_{J/\psi K_S}^{\rm
mix}= -\left(14.1^{+8.1}_{-10.2} \right) \times 10^{-4} \;.
\label{pre}
\end{eqnarray}
Including the CP violation from the $K$-$\bar K$ mixing
\cite{GKL02}, $\Delta S_{J/\psi K_S}$ and $A_{J/\psi K_S}$ remain
$O(10^{-3})$. Equation~(\ref{pre}) will provide an important
standard-model reference for verifying new physics from the
$B^0\to J/\psi K_S$ data in the future. The estimate of $\Delta
S_{J/\psi K_S}\sim O(10^{-4})$ in \cite{HB04} implies that they
have overlooked the more crucial contribution from the penguin
operators. Our observation of the direct CP asymmetry supports the
claim in \cite{HNS06} that $A_{CP}(B\to J/\psi K)$ observed at 1\%
or higher would indicate new physics definitely.

\vskip 1.0cm

We thank W.S. Hou for a useful discussion. This work was supported
by the National Science Council of R.O.C. under Grant No.
NSC-95-2112-M-050-MY3, by the National Center for Theoretical
Sciences, and by the U.S. Department of Energy under Grant No.
DE-FG02-90ER40542. HNL thanks the Yukawa Institute for Theoretical
Physics, Kyoto University, for her hospitality during his visit.

\end{document}